\documentclass[%
reprint,
superscriptaddress,
showpacs,preprintnumbers,
amsmath,amssymb,
aps,
]{revtex4-1}
\usepackage{graphicx}
\usepackage{dcolumn}
\usepackage{xfrac}
\usepackage{bm}
\usepackage{color}
\usepackage[colorinlistoftodos]{todonotes} 
\usepackage[export]{adjustbox}

\newcommand{\uB}{$\mu_B$}

\begin{document}

\title{Interplay of Magnetism and Transport in HoBi}

\author{H.-Y.~Yang}
\affiliation{Department of Physics, Boston College, Chestnut Hill, MA 02467, USA}
\author{J.~Gaudet}
\affiliation{Department of Physics and Astronomy, McMaster University, Hamilton, ON, L8S 4M1, Canada}
\author{A.~A.~Aczel}
\affiliation{Neutron Scattering Division, Oak Ridge National Laboratory, Oak Ridge, TN 37831, USA}
\author{D.~E.~Graf}
\affiliation{National High Magnetic Field Lab, Tallahassee, FL 32310, USA}
\author{P.~Blaha}
\affiliation{Institute of Materials Chemistry, Vienna University of Technology, 1060 Vienna, Austria.}
\author{B.~D.~Gaulin}
\affiliation{Department of Physics and Astronomy, McMaster University, Hamilton, ON L8S 4M1, Canada}
\affiliation{Canadian Institute for Advanced Research, 661 University Ave., Toronto, ON M5G 1M1, Canada}
\affiliation{Brockhouse Institute for Materials Research, Hamilton, ON L8S 4M1 Canada}
\author{Fazel~Tafti}
\email{fazel.tafti@bc.edu}
\affiliation{Department of Physics, Boston College, Chestnut Hill, MA 02467, USA}

\date{\today}

\begin{abstract}
We report the observation of an extreme magnetoresistance (XMR) in HoBi with a large magnetic moment from Ho $f$--electrons.
Neutron scattering is used to determine the magnetic wave vectors across several metamagnetic (MM) transitions on the phase diagram of HoBi.
Unlike other magnetic rare-earth monopnictides, the field dependence of resistivity in HoBi is non-monotonic and reveals clear signatures of every metamagnetic transition in the low-temperature and low-field regime, at $T<2$~K and $H<2.3$~T.
The XMR appears at $H>2.3$~T after all the metamagnetic transitions are complete and the system is spin-polarized by the external magnetic field.
The existence of an onset field for XMR and the intimate connection between magnetism and transport in HoBi are unprecedented among the magnetic rare-earth monopnictides.
Therefore, HoBi provides a unique opportunity to understand the electrical transport in magnetic XMR semimetals.

\end{abstract}

\maketitle



Non-magnetic rare-earth monopnictides with a chemical formula RX where R = Y or La and X = As, Sb, and Bi have attracted attention because they exhibit a non-saturating and extremely large magnetoresistance (XMR)~\cite{pavlosiuk_giant_2016,he_distinct_2016,
yu2017magnetoresistance,xu_origin_2017,pavlosiuk_magnetoresistance_2017,yang_extreme_2017,zeng_compensated_2016,
niu_presence_2016,tafti_resistivity_2016,tafti_temperaturefield_2016,lou_evidence_2017,sun_large_2016,kumar_observation_2016}.
A topological to trivial transition is reported in the LaX family, from LaBi to LaAs, with XMR being present on either side of the transition confirming that XMR originates from an electron-hole compensation instead of a topological band structure~\cite{yang_extreme_2017,nummy_measurement_2018}.
Recently, XMR has been reported in a few magnetic rare-earth monopnictides including CeSb~\cite{ye2018extreme}, NdSb~\cite{wakeham2016large,wang_topological_2018}, GdSb and GdBi~\cite{li1996electrical,ye2018extreme,song_extremely_2018} where $f$--electrons provide localized moments.
In these magnetic semimetals, the itinerant $d$/$p$--electrons couple to the localized $f$--electrons through the Ruderman-Kittel-Kasuya-Yosida  (RKKY) interaction~\cite{ruderman_indirect_1954,kasuya_theory_1956} giving rise to antiferromagnetic (AFM) order, field induced metamagnetic (MM) transitions, and rich magnetic phase diagrams~\cite{busch1967magnetic,rossat1977phase,rossat1980specific,tsuchida1967field,rossat1983magnetic,busch1965magnetic,li1997magnetic}.
Despite complex magnetization curves $M(H)$ with multiple MM transitions, the magnetic monopnictides exhibit plain quadratic resistivity curves $\rho(H)$ and an XMR behavior that is indistinguishable from their non-magnetic analogues in the low-temperature regime ($T<$ 2~K) \cite{ye2018extreme,wang_topological_2018}.
From LaSb/LaBi to CeSb, NdSb, and then GdSb/GdBi, the lanthanide becomes progressively more magnetic, but intriguingly no strong response of transport and XMR to magnetism has been observed so far.
In search of such connection between magnetism and transport properties in a magnetic XMR material, we decided to study HoBi where Ho$^{3+}$ ions provide the largest total angular momentum $\mathbf{J}=\mathbf{L}+\mathbf{S}$ among the R$^{3+}$ ions.
Through a combination of magnetization, neutron scattering, and transport experiments, we unveil an intimate relation between the electronic transport and the magnetism of HoBi unlike any previously studied magnetic RX system.
Using neutron diffraction, we reveal a new $(\sfrac{1}{6},\sfrac{1}{6},\sfrac{1}{6})$ ordered state at intermediate fields which strongly affects the resistivity behavior.
The XMR in HoBi no longer follows a plain quadratic curve and appears only after the magnetic field is strong enough to drive the system out of this $(\sfrac{1}{6},\sfrac{1}{6},\sfrac{1}{6})$ phase and into a $(0,0,0)$ spin polarized state.


Single crystals of HoBi were grown using a self-flux method as described in the Supplemental Material~\cite{suppmatt}.
%
%
Resistivity and heat capacity were measured inside a Quantum Design PPMS Dynacool.
%
DC magnetization was measured inside a Quantum Design MPMS3.
Single crystal neutron diffraction was performed on the HB-1A triple-axis spectrometer at the High Flux Isotope Reactor (HFIR) at the Oak Ridge National Laboratory (ORNL).
%
%
%
Density functional theory (DFT) calculations with full-potential linearized augmented plane-wave (LAPW) method were implemented in the WIEN2k code~\cite{blaha_wien2k_2001} using the Perdew-Burke-Ernzerhof (PBE) exchange-correlation potential~\cite{perdew_generalized_1996}, spin-orbit coupling (SOC), and on-site Coulomb repulsion (Hubbard $U$) in a PBE+SOC+U calculation~\cite{anisimov_band_1991} for the correlated $4f$-electrons.
%
%
High-field experiments were performed in a 35~T DC magnet at the MagLab in Tallahassee inside a $^3$He fridge with a base temperature of 0.3~K.


\begin{figure}
\includegraphics[width=0.48\textwidth,center]{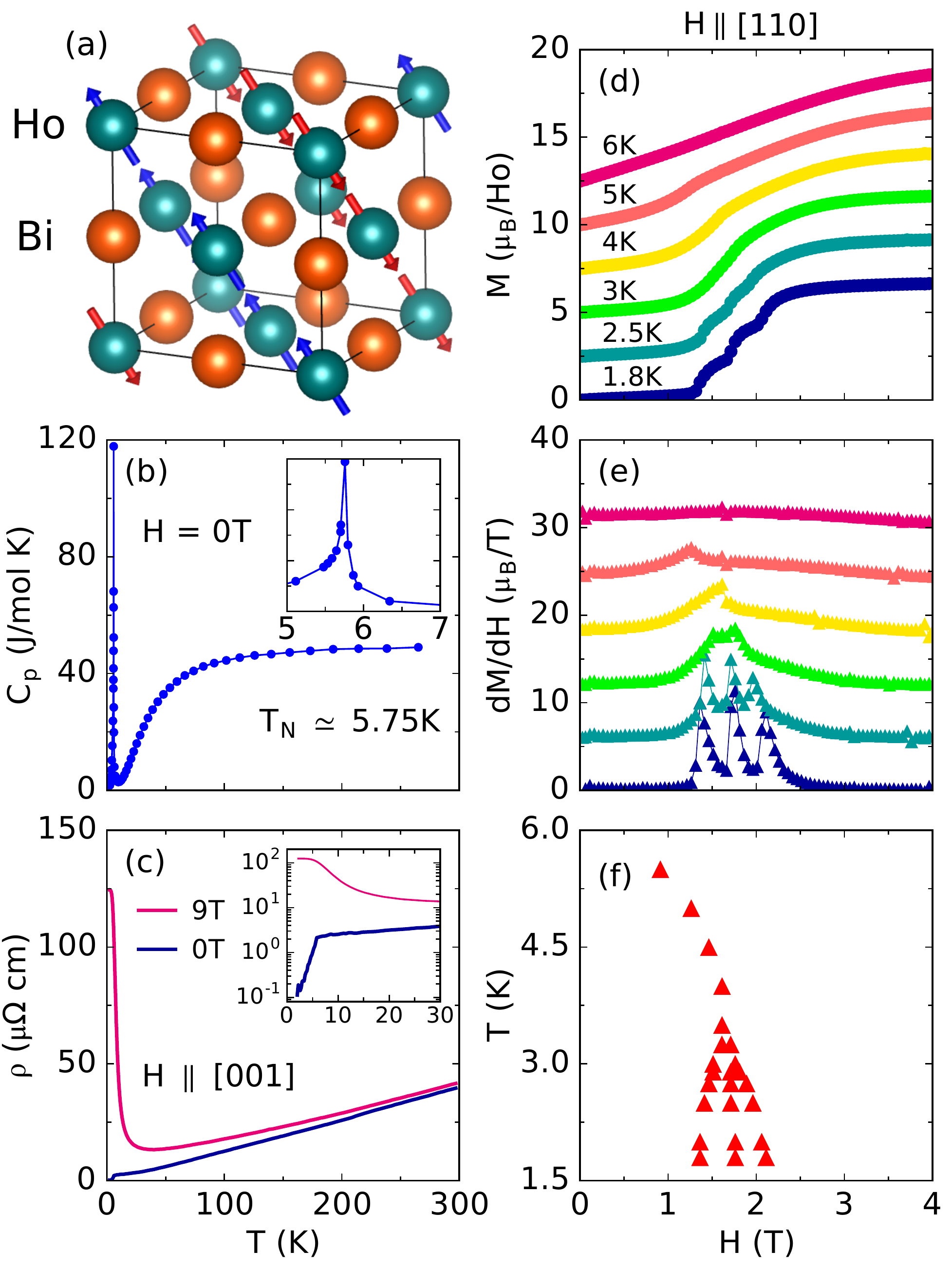}
\caption{\label{XMR}
(a) The rock-salt fcc structure of HoBi ($a=6.23$\AA) with a type-II AFM order.
(b) Heat capacity as a function of temperature at $H=0$ T showing a peak at $T_N=5.75$~K.
(c) Resistivity as a function of temperature at $H=0$ and 9~T showing XMR in a HoBi sample with RRR$(\rho \text{(300K)}/ \rho \text{(0K)})=300$.
(d) Magnetization, in units of \uB~per Ho-atom, as a function of magnetic field ($H\|$[110]) at several representative temperatures.
Curves are shifted for visibility.
(e) $dM/dH$ as a function of field $H\|$[110].
Each peak corresponds to a metamagnetic transition.
(f) Magnetic phase diagram for $H\|$[110] from the peaks in $dM/dH$ data.
}
\end{figure}

Prior studies of HoBi are limited to the magnetization measurements with $H\|$[001] and [111] directions~\cite{hulliger_low_1984,fente_low_2013} as well as neutron scattering in zero-field showing type-II AFM order at $T_N = 5.7$~K~\cite{fischer_magnetic_1985}.
A sketch of the fcc crystal structure of HoBi with the type-II AFM order at $H=0$ is presented in Fig.~\ref{XMR}(a).
Local $f$-moments on Ho-atoms are parallel within each [111] plane and antiparallel between alternate planes.
From the heat capacity measurements in Fig.~\ref{XMR}(b), we confirm the AFM phase transition with a peak at $T_N = 5.75$ K.
From the resistivity measurements in Fig.~\ref{XMR}(c), we reveal a characteristic XMR profile with a large increase of $\rho(T)$ at low temperatures and a resistivity plateau.
The magnitude of XMR is of the order $10^4\%$ when the field is oriented along either [001] or [110] directions (Supplemental Fig.~S1, \cite{suppmatt}).
In this letter, we first extend the magnetic phase diagram of HoBi to the $H\|$[110] direction, which has not been studied before.
Then, we use neutron scattering to determine the magnetic ordering wave vector in each sector of the phase diagram.
Finally, we use the resistivity measurements to show the remarkable connection between the electrical and the magnetic properties of HoBi.
Figure~\ref{XMR}(d) shows the magnetization curves at several representative temperatures with $H\|[110]$.
The steps in the magnetization curves in Fig.~\ref{XMR}(d) correspond to field-induced metamagnetic (MM) transitions which appear as peaks in the $dM/dH$ curves in Fig.~\ref{XMR}(e).
There is only one peak in these data at 3.3~K $<T<T_N$ that marks the boundary of the type-II AFM order.
This single peak splits into three peaks below 3.3~K corresponding to three MM transitions.
A phase diagram is produced in Fig.~\ref{XMR}(f) from the evolution of $dM/dH$ peaks measured at 12 different temperatures between 1.85 and 6~K.
As shown in the Supplemental Material~\cite{suppmatt}, the phase diagram with $H\|[110]$ and [111] have three MM transitions whereas the phase diagram with $H\|[001]$ has six MM transitions.
Here, we focus on the $H\|[110]$ direction because it has a simpler phase diagram compared to the [001] direction and is more accessible to both transport and neutron experiments compared to the [111] direction.

\begin{figure}
\includegraphics[width=0.48\textwidth,center]{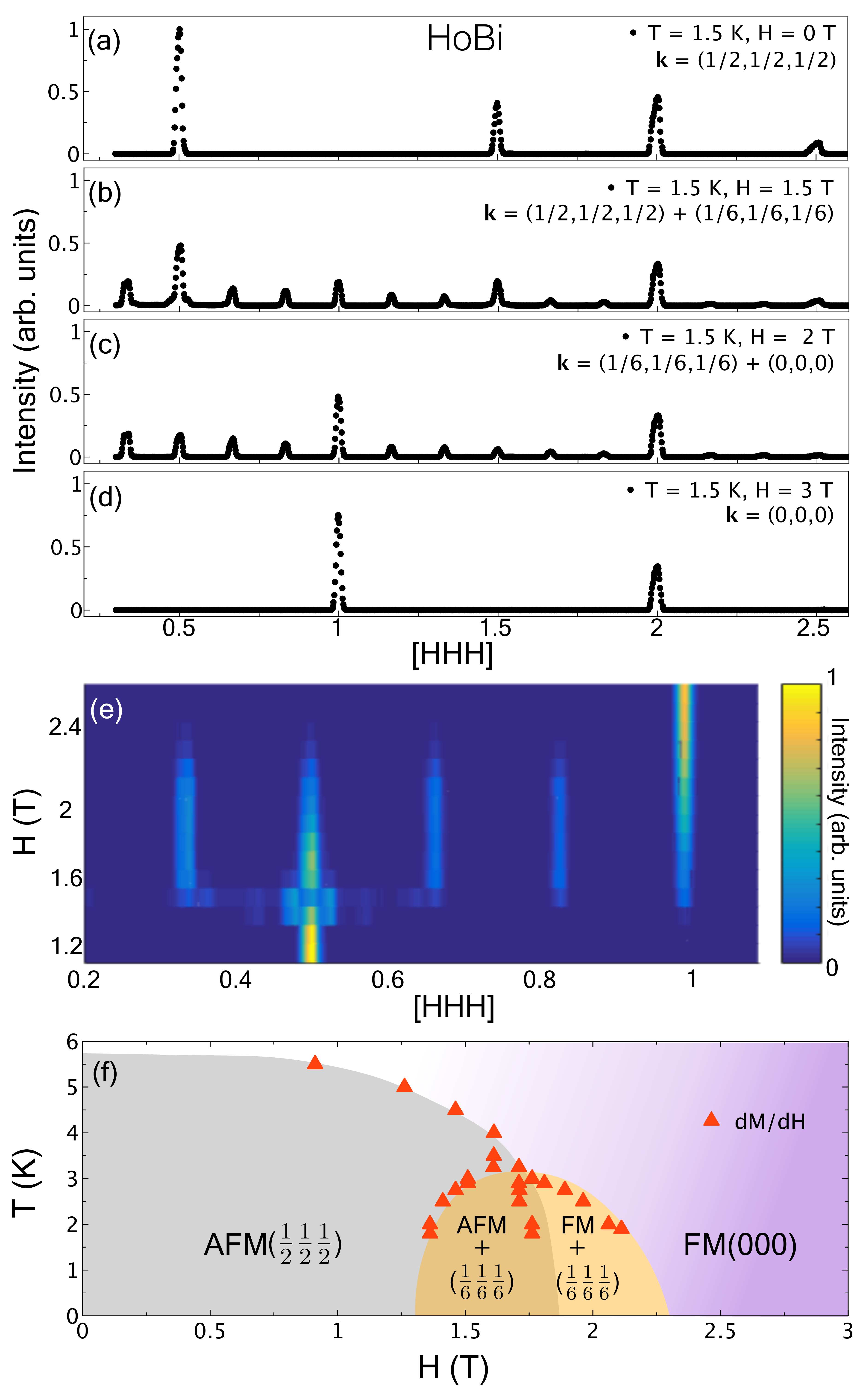}
\caption{\label{NEUTRON}
(a--d) Neutron diffraction intensity along the [HHH] direction at $H=0$,~1.5,~2, and 3~T measured at $T=1.5$~K.
Ordering wave vectors ($\mathbf{k}$) corresponding to each field are listed at the upper-right corner of each panel.
(e) Color map of the neutron diffraction intensity as a function of field and momentum transfer $\mathbf{Q}$ along the [HHH] direction revealing the various metamagnetic transitions in HoBi.
(f) Phase diagram of HoBi according to magnetization and neutron scattering.
}
\end{figure}

To determine the magnetic ordering vector of HoBi in each sector of its phase diagram (Fig.~\ref{XMR}(f)), we turned to neutron scattering.
We performed a broad survey of the neutron diffraction intensity for the momentum transfer $\mathbf{Q}$ that covers the [HHL] plane, perpendicular to the field direction $H\|$[1-10].
This is equivalent to $H\|$[110] used in the magnetization and transport experiments.
Figures~\ref{NEUTRON}(a--d) show representative diffraction patterns along the [HHH] direction at $T=1.5$~K and $H=0$,~1.5,~2,~and 3~T covering all the MM phase transitions.
At each field, structural Bragg peaks appear at $\mathbf{Q}=\mathbf{G}_{hkl}$ with fcc-type Miller indices.
Within each magnetic phase, the Bragg peaks appear at $\mathbf{Q}=\mathbf{G}_{hkl} \pm \mathbf{k}$ where $\mathbf{k}$ is the ordering wave vector.
At $H=0$, Fig.~\ref{NEUTRON}(a) shows magnetic Bragg peaks corresponding to $\mathbf{k}=(\sfrac{1}{2},\sfrac{1}{2},\sfrac{1}{2})$ which specifies the zero-field type-II AFM order below $T_N$.
At $H=1.5$~T, after the first MM transition, new magnetic Bragg peaks appear corresponding to both first and higher order harmonics with $\mathbf{k}=(\sfrac{1}{6},\sfrac{1}{6},\sfrac{1}{6})$.
At $H=2$~T, after the second MM transition, the peak intensities associated with $(\sfrac{1}{6},\sfrac{1}{6},\sfrac{1}{6})$ remain unchanged, the $(\sfrac{1}{2},\sfrac{1}{2},\sfrac{1}{2})$ peak intensities decrease, and a set of $(0,0,0)$ peaks emerge.
The ordering vector $\mathbf{k}=(0,0,0)$ corresponds to a ferromagnetic (FM) alignment of the Ho spins. 
At $H=3$~T, above the third MM transition, the $\mathbf{k}=(\sfrac{1}{6},\sfrac{1}{6},\sfrac{1}{6})$ peaks disappear and the $\mathbf{k}=(0,0,0)$ remains as the only ordering wave vector.

Figure~\ref{NEUTRON}(e) summarizes the results of our measurements at intermediate fields by plotting a color map of the diffraction intensity at different [HHH] vectors.
It shows the appearance of the $(\sfrac{1}{6},\sfrac{1}{6},\sfrac{1}{6})$ order at $1.3<H<2.3$~T, the disappearance of the $(\sfrac{1}{2},\sfrac{1}{2},\sfrac{1}{2})$ AFM order at $H>1.8$~T, and the appearance of $(0,0,0)$ FM order at $H>1.8$~T.
In Fig.~\ref{NEUTRON}(f), a phase diagram of HoBi is constructed based on the magnetization and neutron scattering experiments.
At $T=0$, four distinct phases appear from low to high fields with the ordering wave vectors $\mathbf{k}=(\sfrac{1}{2},\sfrac{1}{2},\sfrac{1}{2})$ (AFM) at $H<1.3$~T, $\mathbf{k}=(\sfrac{1}{2},\sfrac{1}{2},\sfrac{1}{2})$ and $(\sfrac{1}{6},\sfrac{1}{6},\sfrac{1}{6})$ coexisting at $1.3<H<1.8$~T, $\mathbf{k}=(\sfrac{1}{6},\sfrac{1}{6},\sfrac{1}{6})$ and $(0,0,0)$ coexisting at $1.8<H<2.3$~T, and $\mathbf{k}=(0,0,0)$ (FM) at $H>2.3$~T.
At finite temperatures, the $(\sfrac{1}{6},\sfrac{1}{6},\sfrac{1}{6})$ order forms a dome-like boundary at $T<3.3$~K.
The dome is centered around a quantum critical point (QCP) where the AFM $(\sfrac{1}{2},\sfrac{1}{2},\sfrac{1}{2})$ order ends at approximately $H_c=1.8$~T.



\begin{figure}[h]
\includegraphics[width=0.48\textwidth]{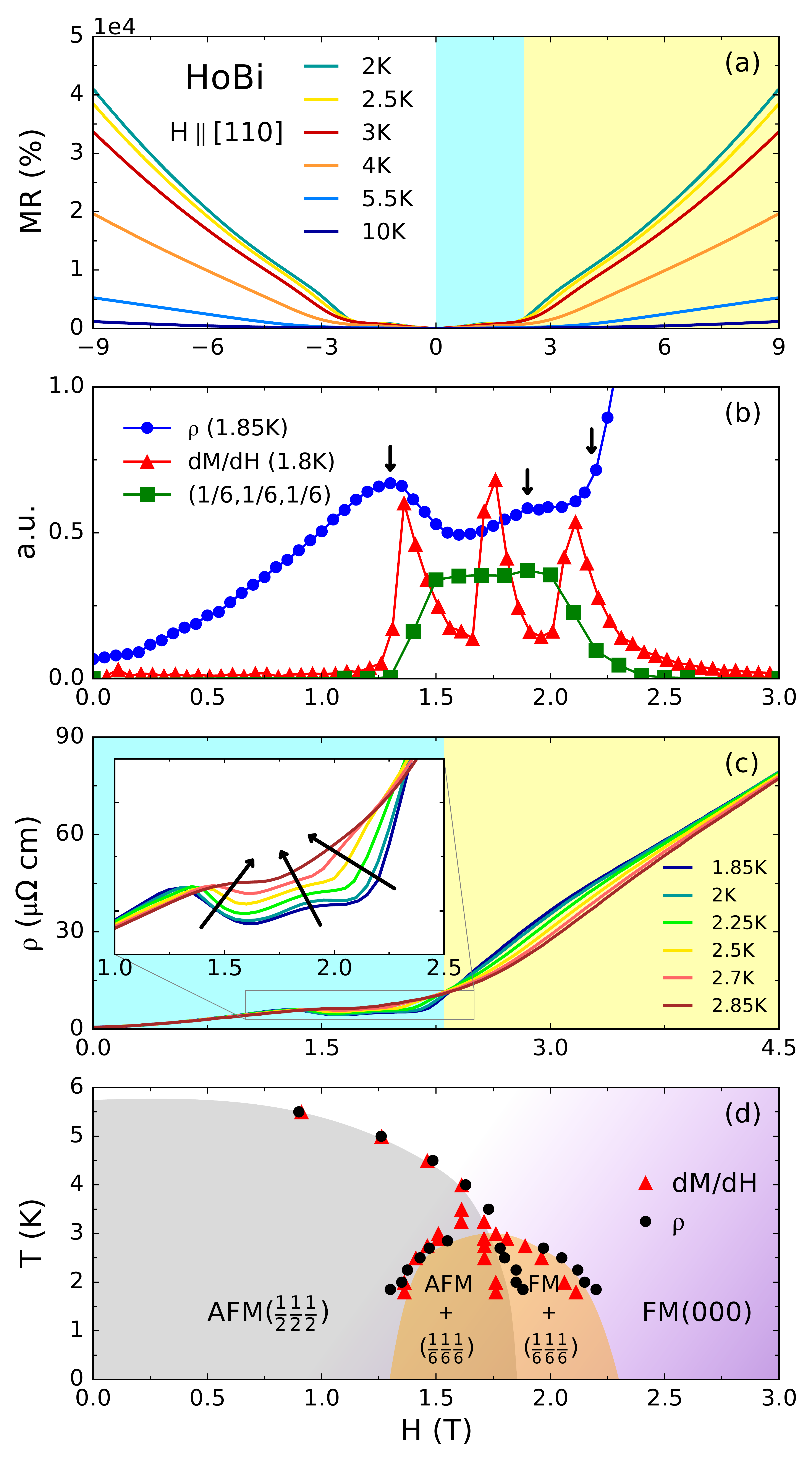}
\caption{\label{DRH}
(a) Magnetoresistance (MR) as a function of field ($H\|$[110], $I\|$[001]) in a HoBi sample with RRR $=127$ at several temperatures.
The region of AFM order and MM transitions is highlighted by blue whereas the region of XMR is yellow.
(b) Resistivity data (blue), $dM/dH$ (red), and the $(\sfrac{1}{6},\sfrac{1}{6},\sfrac{1}{6})$ neutron peak intensity (green) compared at comparable temperatures.
Arrows mark the transport features associated with the MM transitions in $\rho(H)$.
(c) Evolution of $\rho(H)$ with temperature.
Note the onset of XMR (yellow region) at $H=2.3$~T.
(d) Phase diagram of HoBi from the transport, magnetization, and neutron scattering data.
}
\end{figure}

Having established the magnetic phase diagram, we now present the electrical transport data and study the XMR behavior in HoBi.
%
%
%
HoBi shows a typical temperature profile of XMR in Fig.~\ref{XMR}(c) and a large magnitude of MR ($\%$) $=100\times(\rho(H)-\rho(0))/\rho(0)$ in Fig.~\ref{DRH}(a).
What makes HoBi unique among the magnetic monopnictides is an intimate connection between the magnetism and transport that modifies the XMR behavior.
Resistivity measurements in CeSb and NdSb at 2~K show XMR and plain quadratic $\rho(H)$ curves unaffected by the MM transitions~\cite{ye2018extreme,wakeham2016large}.
On the contrary, at 2~K, HoBi exhibits clear features corresponding to each MM transition in the $\rho(H)$ curves.
Furthermore, HoBi has an onset field for XMR at $H=2.3$~T instead of a plain quadratic curve commonly observed in XMR semimetals.

\begin{figure}
\includegraphics[width=0.48\textwidth]{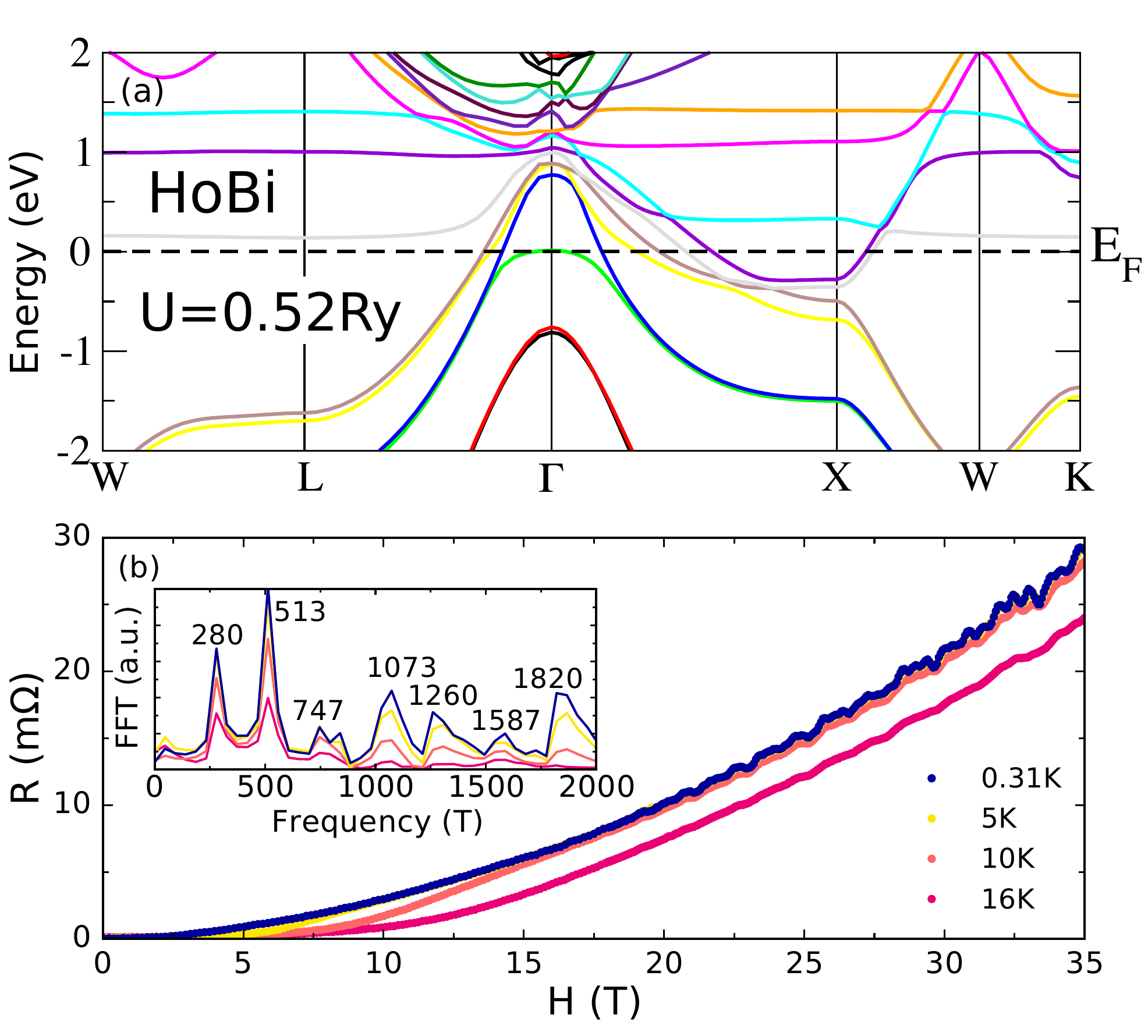}
\caption{\label{QO}
(a) Band structure of HoBi from a spin-polarized PBE+SOC+U calculation with 10,000 $k$--points, basis-size control parameter $RK_{max}=9$, and $U=0.52$~Ry (7.075~eV).
HoBi has four hole bands at $\Gamma$ and two electron bands at $X$ similar to non-magnetic XMR semimetals.
(b) SdH oscillations in electrical resistance at $20<H<35$~T.
Inset shows the Fourier transform of SdH oscillations at several temperatures.
}
\end{figure}

Figure~\ref{DRH}(a) shows two distinct regions in the field dependence of the magnetoresistance MR($H$).
The blue region at $H<2.3$~T is the realm of AFM order and MM transitions.
XMR is absent in this region and does not start until $H>2.3$~T in the yellow region.
Figure~\ref{DRH}(b) compares a representative $\rho(H)$ curve at 1.85~K, a $dM/dH$ curve at 1.8~K, and the intensity of the $(\sfrac{1}{6},\sfrac{1}{6},\sfrac{1}{6})$ neutron diffraction peak at 1.5~K.
With increasing field from zero, $\rho(H)$ shows peaks at the first and second MM transitions, and a steep increase at the third one (see arrows in Fig.~\ref{DRH}(b)).
These features evolve with temperature as shown in the inset of Fig.~\ref{DRH}(c).
Similar features appear in the $\rho(H)$ data at $3.3<T<T_N$ corresponding to the AFM transitions (Supplemental Fig.~S3, \cite{suppmatt}).
The black circles on the phase diagram in Fig.~\ref{DRH}(d) correspond to the AFM and MM transitions derived from the $\rho(H)$ curves.
Remarkably, even without measuring magnetization, one can accurately map the magnetic phase diagram of HoBi using the resistivity data alone.
%

The XMR starts at $H>2.3$~T in the yellow region of Fig.~\ref{DRH}(c).
Such an onset behavior is a unique feature of HoBi and is absent in other rare-earth monopnictides.
At $T=1.85$~K, a steep increase of MR is observed immediately above the onset field 2.3~T followed by a less steep power law behavior at higher fields.
The difference between these two behaviors is better resolved in $d\rho/dH$ curves in Supplemental Fig.~S4~\cite{suppmatt}.
The initial steep MR starts after the $(\sfrac{1}{6},\sfrac{1}{6},\sfrac{1}{6})$ ordering wave vector has disappeared and while the Ho-spins are gradually polarizing with the field to adopt a FM $(0,0,0)$ state.
This is in agreement with the magnetization curve at 1.85~K in Fig.~\ref{XMR}(d) that does not fully saturate until about 3.5~T.
Similarly, the FM $(0,0,0)$ neutron peak intensity keeps increasing with the field until 3.5~T (Supplemental Fig.~S4~\cite{suppmatt}).
Therefore, the disappearance of MM transitions and the gradual polarization of the Ho-spins with field are responsible for the onset of XMR.
Prior studies of non-magnetic monopnictides including LaAs, LaSb, and LaBi have shown a characteristic compensated band structure for XMR with hole pockets at $\Gamma$ and electron pockets at $X$ in the fcc Brillouin zone~\cite{yang_extreme_2017,sun_large_2016}.
We used a combination of DFT calculations and Shubnikov-de Haas (SdH) oscillations to search for such Fermi surfaces in HoBi.
Figure~\ref{QO}(a) shows the calculated band strucutre of HoBi in the XMR (FM) region, which resembles that of non-magnetic XMR materials mentioned above.
There are four hole bands at $\Gamma$ and two electron bands at $X$.
In particular, notice the smallest hole-pocket that barely touches the Fermi level at $\Gamma$.
The size of this pocket is extremely sensitive to the choice of $U$ and a stringent test of the calculation (Supplemental Fig.~S5~\cite{suppmatt}).
%
%
We use the SKEAF program~\cite{rourke_numerical_2012} to calculate the extremal orbit area and the effective mass of carriers on each calculated Fermi surface.
The results are then compared to the frequencies of SdH oscillations in Fig.~\ref{QO}(b).
Fourier transform of the oscillations is shown in the inset with a low-frequency peak at 280~T corresponding to the smallest hole-pocket in Fig.~\ref{QO}(a) with a calculated frequency of 243~T.
Furthermore, we measured the mass of carriers on each pocket using a standard Lifshitz-Kosevich analysis~\cite{shoenberg_magnetic_2009,willardson1967physics} as elaborated in the Supplemental Material~\cite{suppmatt}.
Table~\ref{T1} shows a good agreement between the frequencies and effective masses from DFT calculations and SdH oscillations.
Our analysis of Fermi surfaces confirms that the non-saturating XMR in HoBi originates from a compensated band structure with hole pockets at $\Gamma$ and electron pockets at $X$, similar to the non-magnetic analogue LaBi~\cite{yang_extreme_2017,tafti_temperaturefield_2016,sun_large_2016}.
\begin{table*}
\caption{\label{T1}%
Calculated frequencies and effective masses from the PBE+SOC+U calculations compared to the SdH experimental results in HoBi.
Four hole bands ($h_{1-4}$) and two electron bands ($e_{1,2}$) are observed.
The frequencies $F$ and the effective masses $m^*$ are reported in units of T and $m_e$.
DFT calculations suggest a maximum and a minimum frequency in $h_4$, $e_1$, and $e_2$ .
}
\begin{ruledtabular}
\begin{tabular}{l|c|c|c|c|c|c }
band       & $h_1$ & $h_2$ & $h_3$ & $h_4$ & $e_1$ & $e_2$   \\
           & $F$, $m^*$ & $F$, $m^*$ & $F$, $m^*$ & $F$, $m^*$ & $F$, $m^*$ & $F$, $m^*$ \\
\colrule
DFT        & 243, 1.14  &  1003, 0.20  &  1558, 0.45 &  1975/1986, 0.4/0.42 & 838/1290, 0.47/0.82 & 489/835, 0.54/0.46 \\
\hline
SdH        & 280, 0.27  &  1073, 0.50   &  1587, 0.38 &  1820, 0.58 &  747/1260, 0.29/0.50 &  513/747, 0.29/0.29
\end{tabular}
\end{ruledtabular}
\end{table*}

To summarize, HoBi is the only magnetic rare-earth monopnictide with XMR where the transport behavior, especially XMR, is strongly affected by changes in the magnetic wave vector.
Metamagnetic transitions are resolved in the $\rho(H)$ data clearly so the magnetic phase diagram of HoBi can be accurately mapped from the transport data.
The $(\sfrac{1}{6},\sfrac{1}{6},\sfrac{1}{6})$ dome is intriguing; it affects the XMR behavior drastically and drives its field dependence away from a plain quadratic curve.
%
%
It is likely that the $(\sfrac{1}{6},\sfrac{1}{6},\sfrac{1}{6})$ order is produced by a reconstruction of the Fermi surface at the QCP as a result of special nesting conditions.
It would be interesting to confirm this idea and to search for its consequences such as charge ordering in HoBi.


\section*{ACKNOWLEDGMENTS}

The work at Boston College was funded by the National Science Foundation, Award No. DMR-1708929.
The work at McMater University was funded by NSERC of Canada.
The National High Magnetic Field Laboratory is supported by National Science Foundation through NSF/DMR-1644779 and the State of Florida.
A portion of this work used resources at the High Flux Isotope Reactor, a DOE Office of Science User Facility operated by Oak Ridge National Laboratory.





\bibliography{Yang_HoBi_30may2018}

\end{document}